\documentclass[10pt,journal,compsoc]{IEEEtran}

\usepackage{amsmath,amssymb,amsfonts}
\usepackage{algorithm}
\usepackage{algorithmic}
\usepackage{array}
\usepackage{booktabs}
\usepackage{cite}
\usepackage{graphicx}
\usepackage{stfloats}
\usepackage{textcomp}
\usepackage{url}

\title{DeCoRAG: Cognitive Decoupling and Semantic-Aware Cropping for Complex Document Understanding}

\author{Shuo~Wang, Kai~Zhang, Wenyuan~Huang, Yizheng~Yu, Xia~Liao, Junming~Su, Qing~Wang, and Fang~Xi%
\thanks{\raggedright Corresponding author: F.~Xi.}%
\thanks{\raggedright S.~Wang, K.~Zhang, W.~Huang, Y.~Yu, J.~Su, Q.~Wang, and F.~Xi are with QiYuanLab, Beijing, China. \protect\\
E-mail: \mbox{wangshuo@qiyuanlab.com}; \mbox{zhangkai@qiyuanlab.com}; \mbox{huangwenyuan@qiyuanlab.com}; \mbox{yuyizheng@qiyuanlab.com}; \mbox{sujunming@qiyuanlab.com}; \mbox{d202510468@xs.ustb.edu.cn}; \mbox{xifang@qiyuanlab.com}.}%
\thanks{\raggedright X.~Liao is with Beijing University of Posts and Telecommunications, Beijing, China. \protect\\
E-mail: \mbox{liaoxia@bupt.edu.cn}.}}

\begin{document}

\maketitle

\begin{abstract}
Advancing multimodal retrieval-augmented generation (RAG) for complex document understanding presents a formidable dual dilemma of accuracy and efficiency, particularly in graph RAG. Processing structurally sparse yet visually dense layouts---such as extracting a tiny data marker from a financial chart---often incurs computationally prohibitive token overhead while still triggering catastrophic hallucination. However, multimodal Graph RAG pipelines rely on graph-construction stages that assume Vision-Language Models (VLMs) can resolve sparse semantics within high-density layouts. We challenge this assumption, revealing that forcing VLMs to localize visual evidence, interpret semantics, and extract relations triggers a ``Visual Attention Sink''---a mechanism driving catastrophic semantic loss---while full-page processing incurs massive computational overhead. Controlled interventions verify that this failure is boundary-driven rather than content-specific and that semantic anchoring mitigates it. To fundamentally correct this flawed paradigm, we introduce DeCoRAG, a multimodal Graph RAG pipeline that shifts knowledge processing from coupled visual-semantic reasoning to ``Cognitive Decoupling.'' Rather than passively processing raw pixels, its graph-construction stage establishes a macroscopic Semantic Anchor to neutralize the attention sink. This anchor subsequently drives our Region-Aware Pruning and Cropping (RAP-Crop) mechanism, shifting the reasoning space from dense, noisy backgrounds to purified, intent-driven semantic clusters. The resulting graph supports hybrid retrieval and answer generation. Evaluated on complex document benchmarks, DeCoRAG outperforms dense visual retrievers and multimodal Graph RAG baselines, yielding up to a 12.5\% absolute semantic pass rate improvement over the strongest baseline. DocVQA results demonstrate generalization beyond sparse charts and tables. Crucially, RAP-Crop reduces offline graph-construction prompt tokens by 40.8\% without sacrificing end-to-end accuracy.
\end{abstract}



\section{Introduction}
\label{sec:intro}

The rapid evolution of Large Language Models (LLMs) \cite{touvron2023llama, openai2023gpt4} has established Retrieval-Augmented Generation (RAG) \cite{lewis2020retrieval, guu2020realm, gao2023retrieval, karpukhin2020dense, izacard2021leveraging, guo2025raganything} as the standard paradigm for open-domain QA. To further support complex reasoning, this paradigm is evolving into \textbf{Graph RAG} \cite{edge2024from, he2024g}, which incorporates structured relational representations (e.g., Knowledge Graphs). However, extending Graph RAG to \textbf{high-density industrial documents}---such as financial reports, engineering schematics, and academic papers \cite{huang2022layoutlmv3, kim2022ocr, blecher2023nougat, mathew2021docvqa, singh2019towards}---presents a formidable challenge. Unlike plain web texts, these documents are characterized by a dense interplay of visual charts, tabular data, and unstructured layouts. Extracting high-fidelity structured knowledge from such heterogeneous sources requires not only recognizing optical characters but also precisely comprehending the topological relationships between visual entities.

Current Multimodal \textbf{Graph RAG} systems typically rely on a two-stage pipeline: graph construction (extracting entities and relations) followed by graph-grounded reasoning. However, when applied to complex documents, this pipeline faces a ``Dual Dilemma'' of accuracy and efficiency.
\textbf{First, on the accuracy front (Graph Construction)}, we observe the phenomenon of \textit{Information Collapse}: when existing VLMs \cite{liu2023visual, li2023blip, alayrac2022flamingo, dai2023instructblip} are tasked with extracting \textbf{structured knowledge triples} (e.g., $\langle \textit{Revenue}, \textit{Growth}, \textit{12\%} \rangle$) from complex figures and tables \cite{zhu2023llms}, they often hallucinate \cite{leng2023mitigating, yin2023woodpecker, guan2023hallusionbench} or output sparse relations \cite{li2023evaluating}, yielding incomplete graphs that severely degrade downstream reasoning.
\textbf{Second, on the efficiency front (Visual Encoding)}, addressing the resolution bottleneck creates a prohibitive computational cost. To discern fine-grained details (e.g., axis labels or footnotes), models require high-resolution inputs, yet the visual attention mechanism scales quadratically with image size \cite{dosovitskiy2020image, liu2021swin, dao2022flashattention}. Standard solutions are ill-suited here: sliding windows sever the global semantic topology required for cross-region relations, while downsampling leads to unreadable content.

We argue that this dual dilemma originates from a system-level limitation in current multimodal graph-construction pipelines. Existing systems commonly couple visual localization, semantic interpretation, and relational graph extraction within a single VLM inference process. While operationally simple, this monolithic stage must discover sparse evidence regions, preserve global layout semantics, and induce structured knowledge relations simultaneously. These responsibilities compete for the same visual-token and attention budget, making graph construction both unreliable and inefficient as document resolution and layout complexity increase.
\textbf{First, the accuracy failure manifests as the Visual Attention Sink.} By visualizing the attention distribution, we discover that without explicit semantic guidance, VLMs fail to ground their attention on information-dense visual regions. Instead, the probability mass ``escapes'' to the visual sink tokens (e.g., the top-left corner \cite{darcet2024vision}), mirroring the \textbf{``Attention Sink''} phenomenon observed in streaming LLMs \cite{xiao2024streamingllm, han2023lm, zhang2023h2o}. This model-level failure mode is systematically exposed by the overloaded graph-construction stage, leading to a ``cognitive abandonment'' where the model generates hallucinated graph relations based on language priors rather than visual evidence.
\textbf{Second, the efficiency bottleneck arises because} full-page processing contradicts the principle of \textbf{Visual Sparsity}. Semantically valid information (e.g., discrete data points and textual labels) is often spatially localized, yet full-page high-resolution parsing compels the model to expend a disproportionate computational budget on encoding redundant background regions.
These observations suggest that robust complex-document Graph RAG requires reorganizing the knowledge-construction pipeline rather than merely increasing VLM capacity. They naturally motivate our decoupled design: utilizing global semantic grounding to prevent attention drift, and localized high-resolution extraction to exploit visual sparsity efficiently.

To tackle these challenges, we propose \textbf{DeCoRAG}, a mechanism-guided multimodal Graph RAG system that reorganizes graph construction through a \textbf{``Cognitive Decoupling''} paradigm. 
Departing from the conventional ``end-to-end'' extraction approach, we reformulate the pipeline into three hierarchical phases: \textit{Phase I: Global Semantic Grounding}, \textit{Phase II: Region-Aware Pruning and Cropping (RAP-Crop)}, and \textit{Phase III: Local Structural Extraction}.
In Phase I, we generate a global description that serves as a \textbf{Semantic Anchor}. This anchor performs two critical functions:
(1) \textbf{Attention Redirection:} It acts as a navigational query, forcing the model's attention heads to shift from the sink tokens back to semantically corresponding visual entities, effectively mitigating information collapse.
(2) \textbf{Region-Aware Pruning and Cropping:} Guided by the anchor, we introduce the \textbf{RAP-Crop} algorithm. Adopting a coarse-to-fine strategy, it dynamically identifies and crops target clusters (Clusters-of-Interest), discarding background noise to overcome visual sparsity and maximize effective information density.
Subsequently, these refined regions are fed into Phase III for \textbf{Local Structural Extraction}, ensuring high-fidelity reasoning.

Our contributions are summarized as follows:
\begin{itemize}
    \item \textbf{System-Level Mechanistic Insight:} We identify the ``Visual Attention Sink'' as a model-level failure mode systematically exposed by tightly coupled multimodal graph construction. We substantiate the diagnosis through attention visualization, relation-quality analysis, and controlled paired interventions that separate boundary-driven attention collapse from content-specific artifacts.
    \item \textbf{Mechanism-Guided System Design \& Efficiency:} We propose \textbf{DeCoRAG}, an end-to-end multimodal Graph RAG system built on the \textbf{Cognitive Decoupling} paradigm, which adopts a ``\textbf{Describe-then-Extract}'' approach to bridge macroscopic semantic anchoring with fine-grained evidence grounding. Under this paradigm, our \textbf{RAP-Crop} algorithm dynamically isolates infor\-mation-dense regions, achieving a \textbf{40.8\% prompt-token reduction} and a \textbf{29.6\% total-token reduction} during offline FetaTab graph construction without degrading end-to-end answer quality.
    \item \textbf{Comprehensive System Evaluation:} Extensive experiments on SPIQA, SlideVQA, PaperTab, FetaTab, and DocVQA evaluate end-to-end reliability, component effectiveness, cross-layout generalization, graph-construction efficiency, and robustness. DeCoRAG outperforms state-of-the-art visual retrievers (e.g., \textbf{ColPali} \cite{faysse2024colpali}), multimodal graph frameworks (e.g., \textbf{MMGraphRAG} \cite{wan2025mmgraphrag}), and standard text-only pipelines, while evaluator calibration confirms that the end-to-end Semantic Pass Rate aligns closely with human judgment.
\end{itemize}

\section{Related Work}
\label{sec:related_work}

\textbf{Multimodal and Graph-Based RAG.} 
Retrieval-Augmented Generation (RAG) \cite{lewis2020retrieval} has greatly advanced the knowledge-intensive reasoning capabilities of LLMs. To handle complex relational reasoning, Graph RAG incorporates structured Knowledge Graphs into the retrieval pipeline. Concurrently, to process visually rich documents, multimodal retrievers like ColPali \cite{faysse2024colpali} and VisRAG \cite{hu2024visrag} leverage VLMs for direct patch-level matching. Recent pioneering works, such as MegaRAG \cite{hsiao2025megarag} and MMGraphRAG \cite{wan2025mmgraphrag}, attempt to bridge these two paradigms by constructing multimodal knowledge graphs that integrate visual and textual entities. However, these methods primarily focus on macro-level retrieval and fusion, often overlooking the microscopic feature collapse that occurs when VLMs directly encode high-density, structurally sparse documents. DeCoRAG addresses this gap through a mechanism-guided ``Describe-then-Extract'' system that decouples semantic planning from high-resolution graph extraction.

\textbf{Attention Sink and VLM Hallucination.} 
The phenomenon of the ``Attention Sink'' was initially identified in streaming LLMs \cite{xiao2024streamingllm}, where attention mass disproportionately accumulates on initial sink tokens rather than semantically meaningful features. A similar vulnerability has been observed in Vision Transformers (ViTs), leading to the introduction of Vision Registers \cite{darcet2024vision} to absorb redundant attention. In multimodal understanding, this lack of focused visual grounding is a primary catalyst for object hallucination \cite{li2023evaluating}. In DeCoRAG, we empirically demonstrate that explicitly anchoring the VLM with a global semantic description successfully redirects attention away from these visual traps, thereby recovering the model's complex structural extraction capabilities.

\textbf{Image Cropping for Visual Efficiency.} 
The processing of high-resolution documents poses a severe computational bottleneck. Traditional image cropping methods typically rely on candidate box generation and ranking, often incorporating spatial-aware features and rank consistency \cite{spatial2023cropping, zeng2019reliable}. Recent advances explore end-to-end regression architectures or utilize VLMs via in-context learning to achieve free-form and subject-aware cropping (e.g., Cropper \cite{li2024cropper}). While these methods excel in photographic composition and aesthetic enhancement, they are suboptimal for document parsing \cite{ye2023mplug}, where the goal is to preserve discrete, structurally sparse semantic elements (e.g., data markers and headers) rather than achieving visual harmony. Our RAP-Crop algorithm departs from aesthetic cropping by dynamically fusing visual density priors with VLM-guided semantic anchors, acting as a targeted token pruning mechanism that maximizes effective information density for complex document reasoning.

\section{Analysis of Coupled Multimodal Graph Construction}
\label{sec:analysis}

\begin{figure*}[!t]
    \centering
    \begin{minipage}{0.48\textwidth}
        \centering
        \textbf{(a) Baseline (Standard Inference)}\\[1mm]
        
        \includegraphics[height=2.8cm, keepaspectratio]{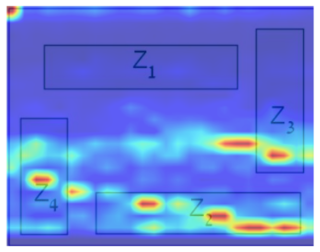}\\[0.5mm]
        {\small \textit{Attention Drift (Top-left Bias)}}\\[2mm]
        
        \includegraphics[width=0.82\linewidth, keepaspectratio]{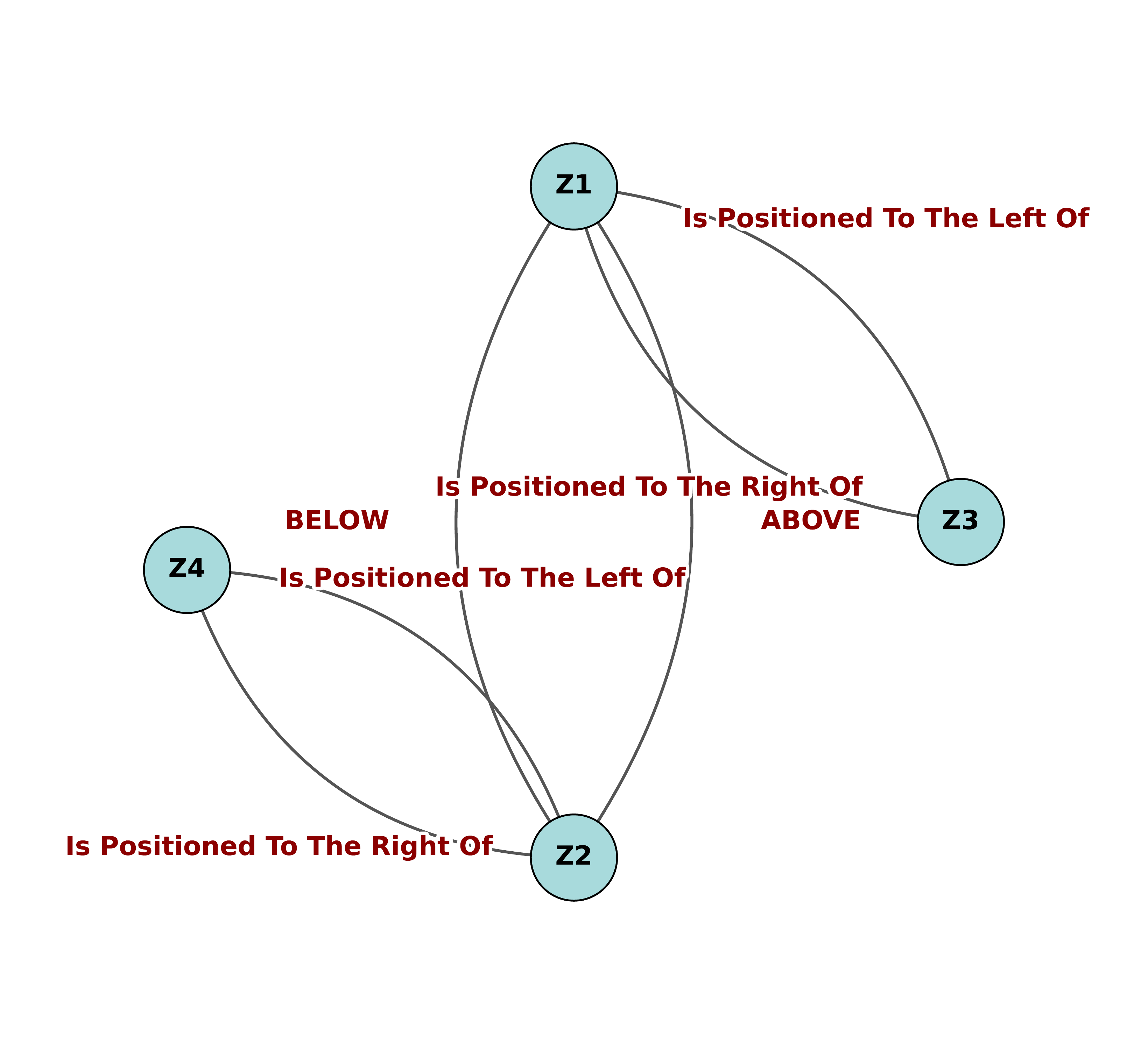}\\[0.5mm]
        {\small \textit{Shallow Spatial Graph}}
    \end{minipage}
    \hfill
    \begin{minipage}{0.48\textwidth}
        \centering
        \textbf{(b) Ours (Visual Grounding)}\\[1mm]
        
        \includegraphics[height=2.8cm, keepaspectratio]{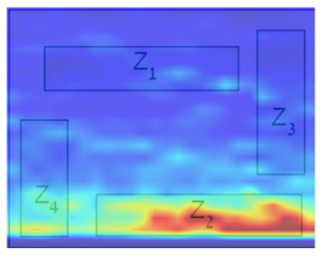}\\[0.5mm]
        {\small \textit{Semantic Anchoring ($Z_2$ Pivot)}}\\[2mm]
        
        \includegraphics[width=0.82\linewidth, keepaspectratio]{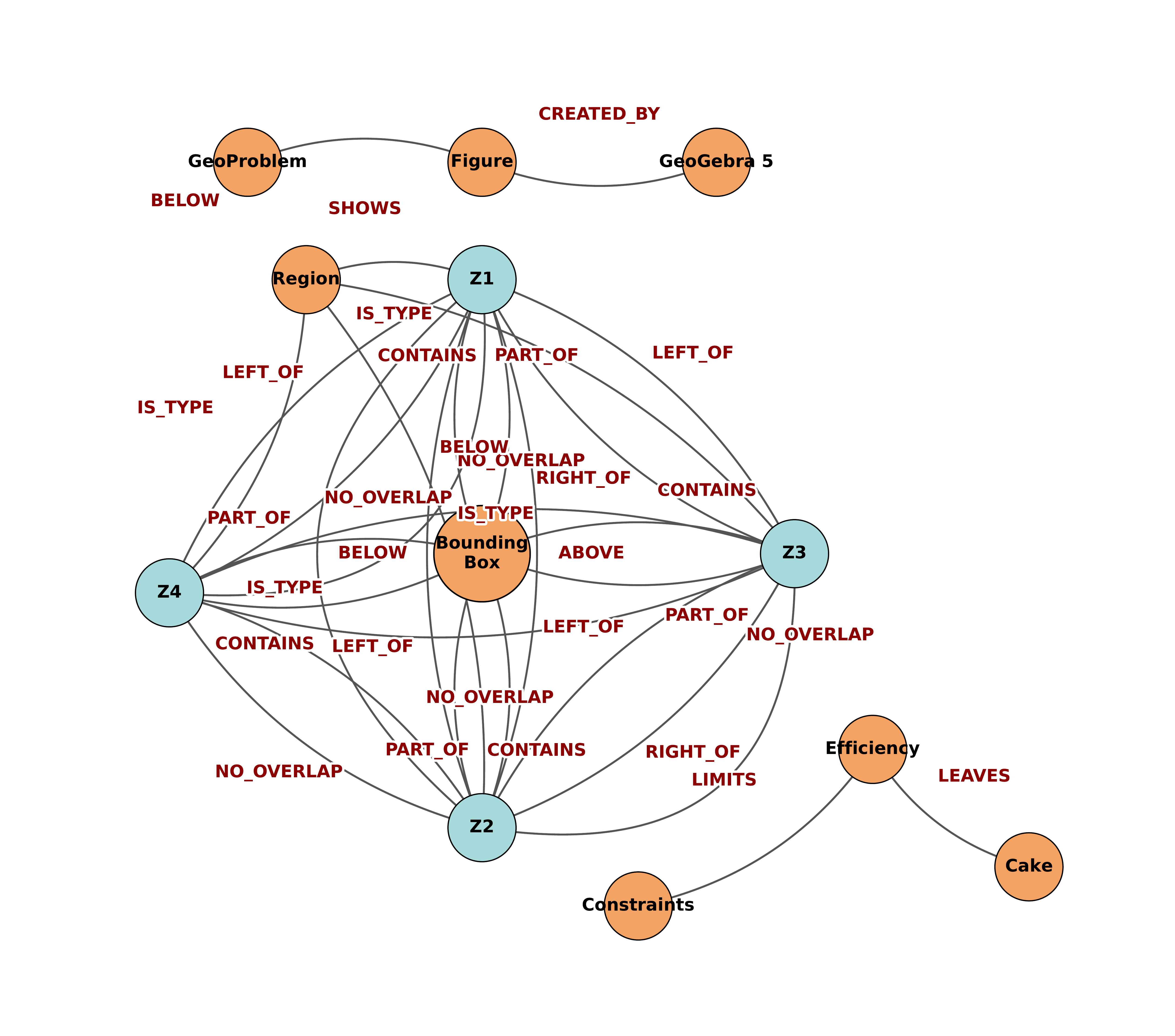}\\[0.5mm]
        {\small \textit{Rich Semantic Graph}}
    \end{minipage}
    
    \caption{\textbf{Comparison of Attention Patterns and Graph Generation.} 
    \textbf{(a) Baseline:} The attention is \textbf{disproportionately attracted to} the top-left corner (``Attention Sink''), despite scattered focus on other regions. This lack of coherent focus results in a \textbf{Shallow Spatial Graph} (bottom), capturing only fragmented positional relations while missing structural logic.
    \textbf{(b) Ours:} Our method effectively anchors attention to semantic regions ($Z_1$-$Z_4$), with a strong focus on $Z_2$ as a \textbf{pivotal spatial anchor} that connects surrounding entities. This precise grounding enables the recovery of a \textbf{Rich Semantic Graph} (bottom) containing complex logical dependencies (e.g., \textit{part\_of, non-overlap}).}
    
    \label{fig:sink_contrast}
\end{figure*}

Before detailing our system design, we analyze why conventional end-to-end multimodal graph-construction pipelines fail on complex documents. These pipelines require a VLM to localize visual evidence, interpret document semantics, and induce graph relations within a single coupled inference process. We empirically show that this coupled formulation is inherently fragile: when sparse semantic elements are embedded in high-resolution pages with large redundant backgrounds, the model often loses reliable visual grounding and produces incomplete or hallucinated graph relations. We identify the \textbf{Visual Attention Sink} as a measurable manifestation of this coupled-pipeline failure, thereby motivating explicit semantic anchoring before fine-grained graph extraction.

\subsection{Experimental Setup: Visualizing Attention Flow}
Unless otherwise specified, all analyses in this section use Qwen-VL \cite{bai2023qwenvl} under the same inference configuration. To reveal the underlying mechanism, we extract the attention weights from the final transformer layer during the autoregressive generation process. 
For each generated text token $t$, we compute a spatial attention map $M_t \in \mathbb{R}^{H \times W}$ by averaging the attention weights across all heads and reshaping the sequence of visual tokens back to the 2D spatial grid of resolution $H \times W$.
To capture the global visual grounding throughout the reasoning phase, we perform temporal aggregation to distill the \textbf{spatial distribution} of attention: $\bar{M} = \frac{1}{L} \sum_{t=1}^{L} M_t$, where $L$ denotes the total length of the generated text sequence. The resulting heatmap $\bar{M}$ quantifies the model's cumulative focus on specific visual regions, intuitively serving as a ``where the model looked'' summary over the entire reasoning process.

\subsection{The Failure Mode: Visual Attention Sink}
\textbf{Task Definition: Open-Schema Visual Extraction.} 
Multimodal Knowledge Graphs have emerged as a promising paradigm for bridging the semantic gap between unstructured pixel inputs and interpretable reasoning \cite{hsiao2025megarag, wan2025mmgraphrag}. A central challenge in this paradigm is to extract structured knowledge from information-intensive visual documents without assuming a pre-defined schema. Inspired by dynamic schema induction \cite{autoschemakg}, we formally define open-schema visual extraction as follows: given a figure or table image $I$, the model must induce a set of triples $\mathcal{G} = \{(h, r, t)\}$ capturing both \textbf{semantic data} (e.g., $\langle \text{Revenue}, \text{increase\_by}, \text{12\%} \rangle$) and \textbf{visual topology} (e.g., $\langle \text{Legend}, \text{below}, \text{X-Axis} \rangle$).

\textbf{Observation: The Failure Mode of Unanchored Inference.} 
To investigate the model's \textbf{inherent behavior} on this complex visual reasoning task, we evaluate the standard inference pipeline, as visualized in Figure \ref{fig:sink_contrast}a.

\textbf{Attention Drift \& Spatial Shallowing.} 
Under this baseline setting, a distinct failure mode emerges. Acting as a \textbf{physical manifestation of high uncertainty} \cite{xiao2024streamingllm}, the attention mass is disproportionately attracted to the top-left corner (``Attention Sink''), failing to focus on specific objects. 
This visual detachment leads to a \textbf{Shallow Spatial Graph}. 
Quantitatively, the extraction yield \textbf{is severely stunted}, capturing only 8 triples (missing over 70\% of the 28 ground-truth relations present in the image).
Qualitatively, the model suffers a severe ``Semantic Loss'': it \textbf{degenerates} complex topological constraints (e.g., \textit{does\_not\_overlap}, \textit{part\_of}) into trivial positional prepositions (e.g., \textit{above}, \textit{left\_of}).
Furthermore, it fails to classify $Z_1$ as a \textit{rectangular region} (\textbf{missing \textit{is\_type} edges}), 
treating it merely as an abstract symbol.

\subsection{The Systemic Trigger: Structural Visual Sparsity}
Unlike natural images characterized by spatially continuous visual semantics, information-intensive visual documents (e.g., charts, tables, and presentation slides) exhibit \textbf{Structural Visual Sparsity}.
As revealed by our empirical investigation, the valid \textbf{``Semantic Elements''} (e.g., textual tokens, geometric primitives, and data markers) are \textbf{discretely distributed} across the image canvas. In contrast, the vast majority of the pixel space is dominated by non-informative background regions (e.g., whitespace, padding, or decorative borders).

This inherent sparsity poses a fundamental computational challenge. Since standard Vision Transformers process images as uniform patches, the model is forced to expend a disproportionate amount of its computational budget on encoding redundant background features. 
Physically, this results in an extremely low \textbf{Effective Information Density}, where the attention mechanism struggles to filter the sparse, localized signals from the overwhelming background, thereby increasing the susceptibility to attention drift.

\textbf{Motivation for Cognitive Decoupling:} Consequently, a graph-construction stage that relies on unguided self-attention to localize sparse evidence and extract complex relations simultaneously is inherently fragile. Structural Visual Sparsity amplifies the model's tendency to lose visual grounding amidst redundant background pixels. We therefore formulate a systematic countermeasure: the pipeline first establishes an explicit cognitive prior---a macroscopic \textit{Semantic Anchor}---before performing fine-grained visual extraction.

\subsection{The Solution: Semantic Anchoring \& Graph Recovery}
\label{subsec:the_cure}

\textbf{The Verification (Ours): Reversing the Collapse.} 
To verify the efficacy of our proposed method in addressing this limitation, we introduce a \textbf{Visual Grounding} intervention (Figure \ref{fig:sink_contrast}b).
By explicitly anchoring attention to semantic regions ($Z_1$-$Z_4$), the extracted graph successfully establishes $Z_2$ as a \textbf{pivotal spatial anchor}, as extracted triples consistently employ it as the \textit{spatial datum} to localize surrounding entities.
This intervention reverses the collapse, recovering a \textbf{Rich Semantic Graph} that retains the full density of 28 triples. 
Crucially, it restores the missing high-level logical dependencies (e.g., \textit{non-overlap} constraints among siblings), showing that stronger visual grounding supports complex logical reasoning.

\begin{figure*}[!t]
    \centering
    \setlength{\tabcolsep}{2pt}
    \begin{tabular}{ccc}
        \includegraphics[height=5.2cm, keepaspectratio]{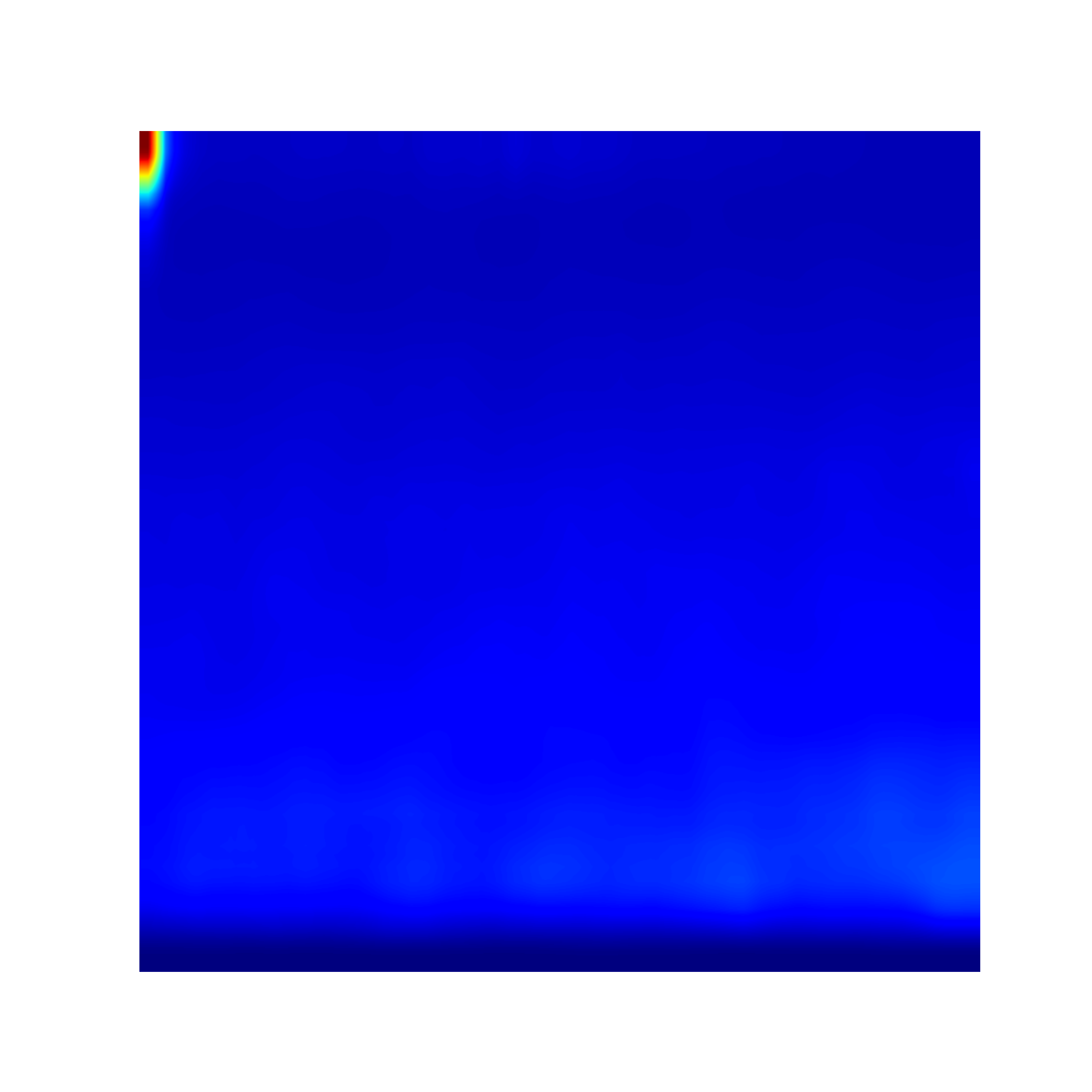} & 
        \includegraphics[height=5.2cm, keepaspectratio]{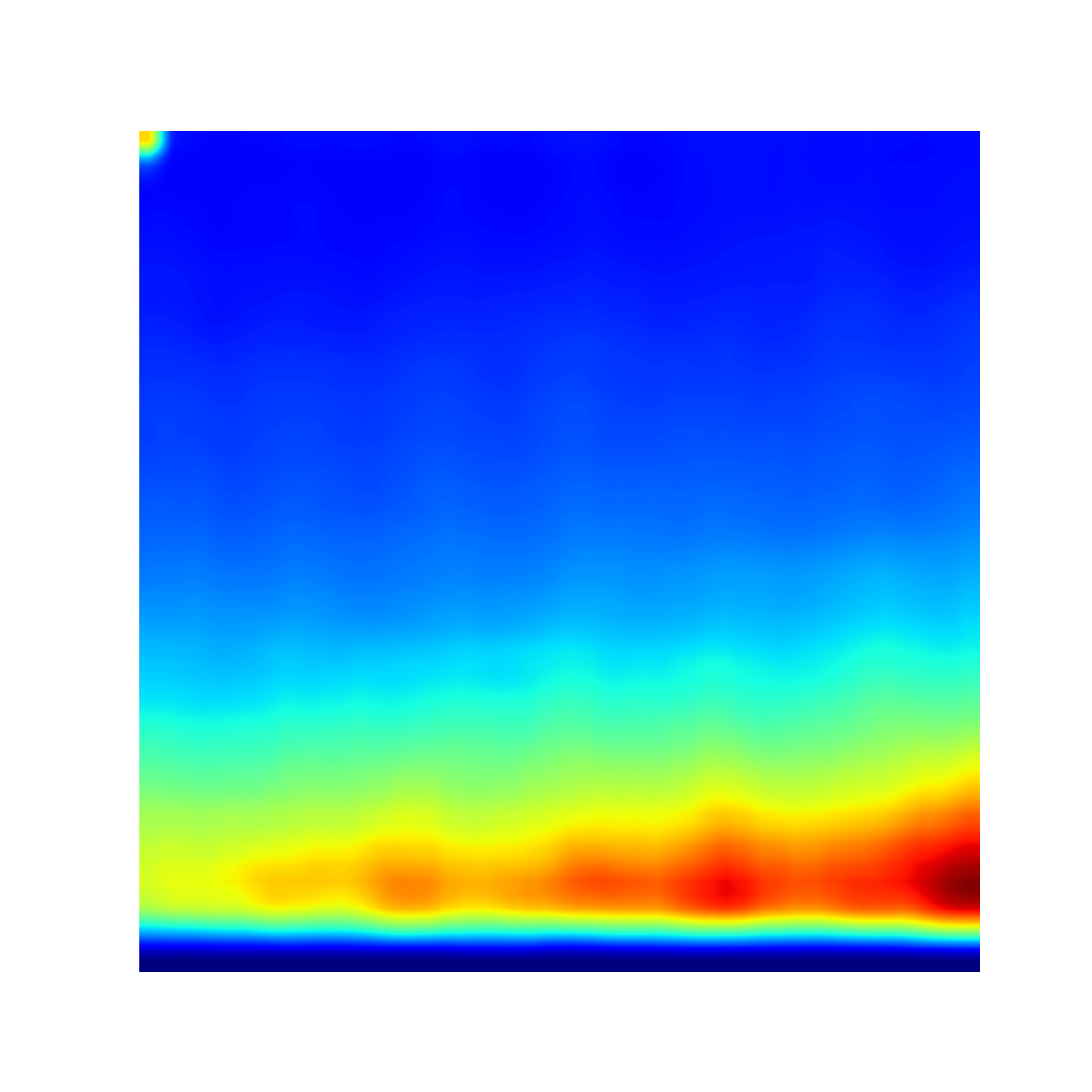} & 
        \includegraphics[height=5.2cm, keepaspectratio]{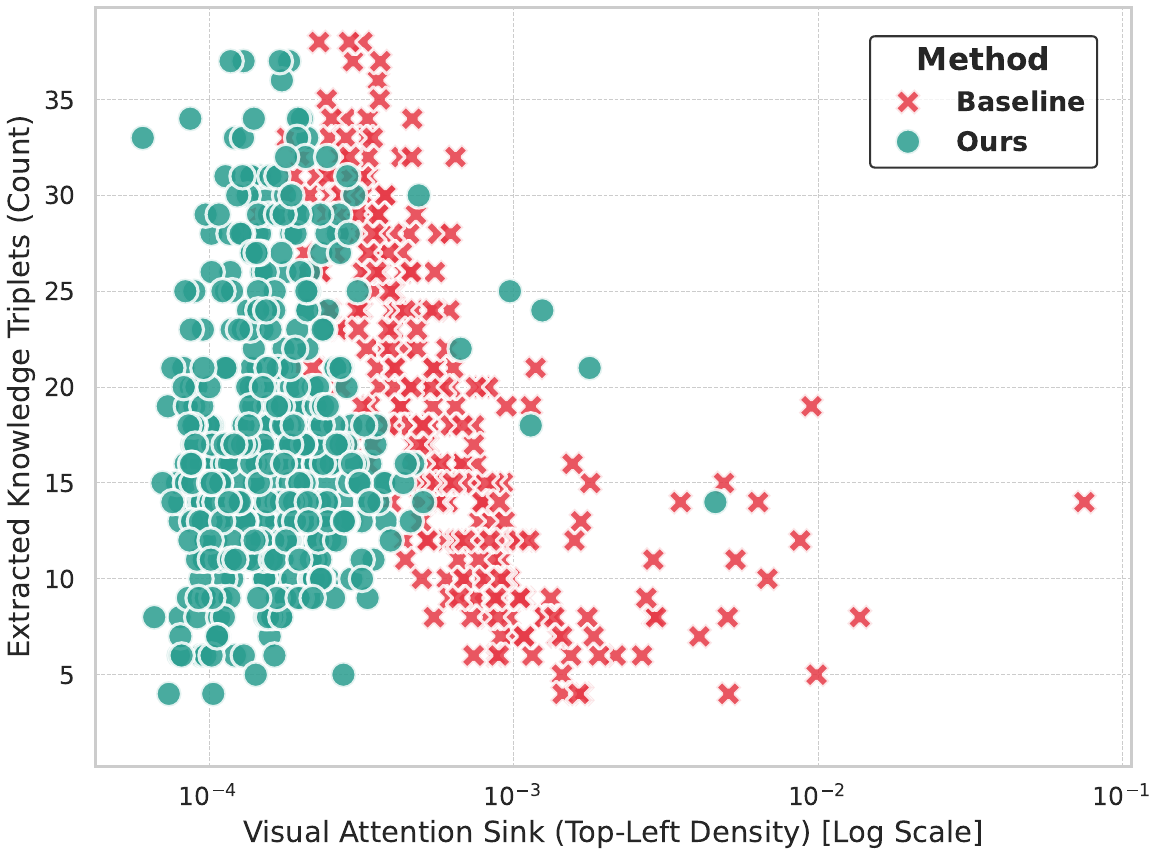} \\
        {\small \textbf{(a) Avg. Map (Baseline)}} & 
        {\small \textbf{(b) Avg. Map (Ours)}} & 
        {\small \textbf{(c) Sink Density vs. Extraction Yield}} \\
    \end{tabular}
    \caption{\textbf{Quantitative Validation of the Attention Sink Phenomenon.} 
    \textbf{(a)} The aggregated attention map over 500 sampled SPIQA images reveals a persistent hotspot at the top-left patch during baseline inference. 
    \textbf{(b)} Our Visual Grounding method suppresses this sink, redistributing attention towards the bottom-center regions. 
    \textbf{(c)} The scatter plot illustrates an observable negative trend between sink severity (X-axis) and extraction yield (Y-axis). Our method effectively shifts the sample distribution leftwards into a low-activation regime, recovering the inherent extraction capacity.}
    \label{fig:sink_statistics}
\end{figure*}

\textbf{Quantitative Evidence: Prevalence and Correlation.} 
To verify that the ``Attention Sink'' is a systemic issue rather than an isolated case, we conduct a quantitative analysis on 500 complex figure and table images randomly sampled from the SPIQA dataset \cite{pramanick2024spiqa}, a comprehensive benchmark featuring highly complex and information-dense visual documents. First, we aggregate the attention maps across all samples to compute the \textbf{Global Average Attention Maps}\footnote{Visualizing raw attention is often hindered by the extreme saturation at the image boundaries (particularly the bottom-right corner), which corresponds to sequence separators or boundary tokens. To prevent this artifactual attention from masking the authentic intra-image spatial attention distribution, we zero out the bottommost row during visualization.}.
Under the baseline setting (Figure \ref{fig:sink_statistics}a), the visualization reveals a pronounced and persistent hotspot strictly localized at the top-left patch, providing definitive visual proof that the model universally defaults to this region. In stark contrast, applying our Visual Grounding intervention (Figure \ref{fig:sink_statistics}b) effectively mitigates this sink, redistributing the attention mass towards the bottom-center regions where core structural elements typically reside.

Furthermore, to establish the causal impact of this phenomenon on extraction performance, we plot the relationship between the severity of the attention sink (Top-Left Density) and the extraction yield (Extracted Knowledge Triplets) in Figure \ref{fig:sink_statistics}c. 
An observable \textbf{negative trend} is evident in the Baseline setting (red crosses): as the model allocates more attention to the non-semantic sink region (shifting rightwards on the log scale), the extraction capability severely deteriorates, with the majority of high-sink samples trapped below 15 triples. 
Conversely, our Visual Grounding intervention (green circles) systematically suppresses this spurious activation, shifting the sink density leftwards into a benign, low-activation regime. By decoupling the attention mechanism from this visual trap, our method successfully restores the model's inherent extraction capacity.

\subsection{Causal Diagnosis Through Controlled Interventions}
\label{subsec:causal_intervention}

The prevalence and negative correlation above establish that the top-left sink is systematic, but correlation alone does not distinguish a boundary-driven attention attractor from a specific visual artifact in the original top-left pixels. We therefore conduct a paired diagnostic study on 100 SPIQA pages using the same Qwen-VL backbone and inference configuration. Each condition preserves the same image-prompt pair as the original input and modifies only the visual boundary, the top-left content, or the semantic prior. We report two sink statistics: the raw top-left attention mass and the Corner Over-Concentration Ratio (COCR), defined over the temporally aggregated attention map $\bar{M}$ as
\begin{equation}
\operatorname{COCR}(\bar{M}) =
\frac{\frac{1}{|\Omega_{\mathrm{TL}}|}\sum_{(i,j)\in \Omega_{\mathrm{TL}}}\bar{M}_{ij}}
{\frac{1}{HW}\sum_{i=1}^{H}\sum_{j=1}^{W}\bar{M}_{ij}},
\label{eq:cocr}
\end{equation}
where $\Omega_{\mathrm{TL}}$ denotes a fixed top-left diagnostic window. A COCR greater than 1 indicates that the top-left window receives more attention density than an average image region.

\begin{table*}[!t]
\caption{\textbf{Controlled Intervention Study for the Visual Attention Sink.} Values are means over 100 paired SPIQA pages; parentheses denote mean paired changes relative to Original. TL/BR denote top-left and bottom-right.}
\label{tab:sink_intervention}
\centering
\footnotesize
\renewcommand{\arraystretch}{1.08}
\setlength{\tabcolsep}{3.4pt}
\begin{tabular*}{\textwidth}{@{\extracolsep{\fill}}p{0.15\textwidth}p{0.22\textwidth}ccp{0.30\textwidth}@{}}
\toprule
\textbf{Input Condition} & \textbf{Causal Test} & \textbf{COCR$\downarrow$} & \textbf{TL Mass$\downarrow$} & \textbf{Paired Finding} \\
\midrule
Original & baseline & 3.49 & 0.0146 & unmodified input \\
TL mask & remove TL pixels & 3.56 (+0.08) & 0.0155 (+0.0008) & no systematic change; not content-driven \\
Control mask & mask non-TL corner & 6.62 (+3.14) & 0.0220 (+0.0074) & generic masking does not mitigate the sink \\
TL blank margin & add blank TL area & 3.56 (+0.08) & 0.0158 (+0.0011) & attention leaves the original content's TL region \\
BR blank margin & canvas-size control & 2.70 (-0.79) & 0.0113 (-0.0033) & weak boundary-only reduction \\
TL shift & alter image layout & 12.57 (+9.09) & 0.0221 (+0.0074) & unstable; attention follows the new boundary \\
Semantic anchor & proposed mitigation & \textbf{0.80 (-2.68)} & \textbf{0.0045 (-0.0102)} & \textbf{88/100 improve; validity 75\%$\to$85\%} \\
\bottomrule
\end{tabular*}
\end{table*}

Table \ref{tab:sink_intervention} provides three mechanistic conclusions. First, masking the original top-left pixels barely changes either COCR or TL mass, which rules out the hypothesis that the hotspot is caused by a specific visible object or OCR fragment located in that region. Second, blank-margin and layout-shift interventions show that unanchored attention is sensitive to image boundaries: the attention mass can leave the original content's top-left area and reappear around the newly induced boundary region. Third, these pixel- and layout-level edits are diagnostic rather than reliable mitigations; generic masking can even amplify over-concentration. In contrast, semantic anchoring directly changes the reasoning prior and robustly suppresses the sink, reducing COCR by 77\% and TL mass by 69\%. The paired improvement is observed in 88/100 samples (sign test $p\!\approx\!1.9{\times}10^{-15}$). The same samples also improve structured-output validity from 75\% to 85\%, indicating that attention redirection is coupled with graph quality rather than being a purely cosmetic heatmap effect.

\textbf{Beyond Quantity: Information Richness and Relational Depth.} 
While the restored extraction yield is clearly evident, a higher triplet count does not inherently guarantee richer semantics, as unguided models often inflate outputs with hallucinated entities or trivial spatial prepositions. To rigorously quantify this on the same sampled dataset, we classify the generated relations into a four-tier hierarchy: \textbf{L0 (Hallucination/Invalid)}, \textbf{L1 (Trivial Spatial)} (e.g., \textit{above}), \textbf{L2 (Basic Attribute)} (e.g., \textit{has}), and \textbf{L3 (High-Order Logical)} (e.g., \textit{increase\_by}). 

\begin{figure}[!t]
    \centering
    \includegraphics[width=0.9\linewidth]{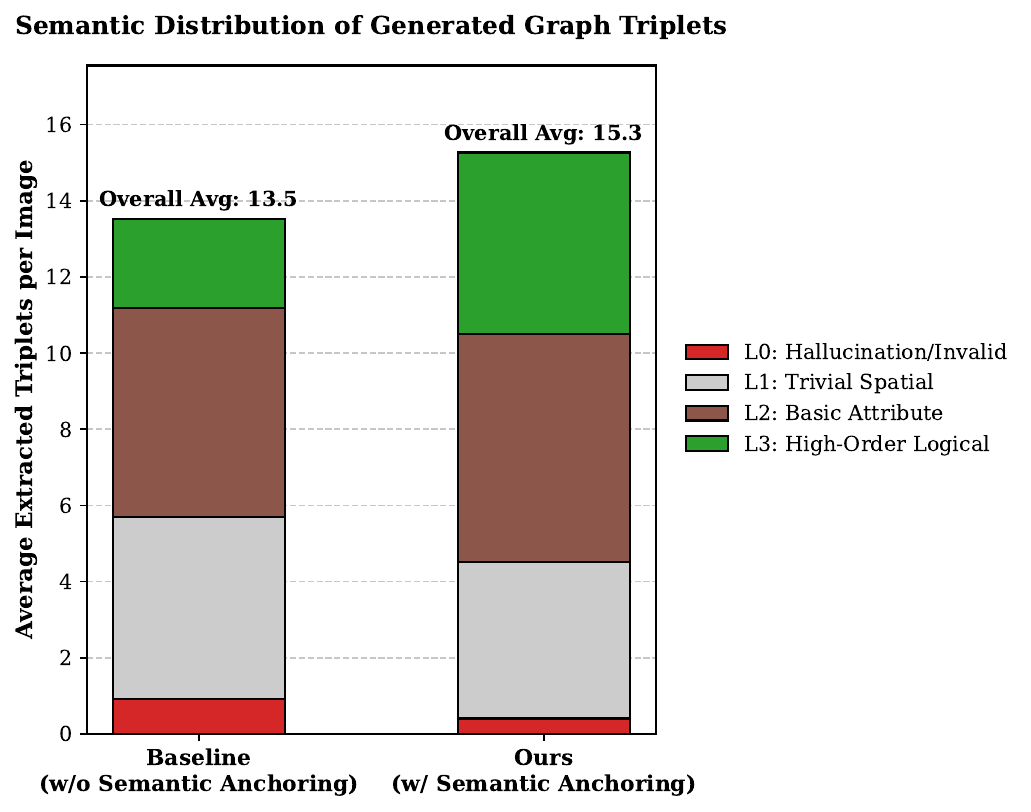}
    \caption{\textbf{Semantic Distribution of Generated Triplets.} Explicit Semantic Anchoring effectively suppresses L0 and L1 noise while significantly expanding the density of high-order logical triples.}
    \label{fig:semantic_distribution}
\end{figure}

\begin{table}[!t]
    \centering
    \footnotesize
    \caption{\textbf{Information Richness Assessment.} Noise Ratio = (L0+L1)/Total. DSSR measures core L3 fact retention.}
    \label{tab:information_richness}
    \setlength{\tabcolsep}{3.5pt}
    \begin{tabular}{lcccc}
    \toprule
    \textbf{Method} & \textbf{Avg. Trip.} & \textbf{Noise $\downarrow$} & \textbf{RDI $\uparrow$} & \textbf{DSSR $\uparrow$} \\
    \midrule
    w/o Anchoring & 13.5 & 42.1\% & 57.3\% & - \\
    \textbf{w/ Anchoring} & \textbf{15.3} & \textbf{29.5\%} & \textbf{70.5\%} & \textbf{84.0\%} \\
    \bottomrule
    \end{tabular}
\end{table}

As detailed in Table \ref{tab:information_richness} and Figure \ref{fig:semantic_distribution}, without semantic anchoring (Baseline), the graph generation is severely inflated by trivial spatial relations and visual hallucinations (Noise Ratio: 42.1\%). In stark contrast, explicit anchoring substantially suppresses this noise and boosts the \textbf{Relational Depth Index (RDI)}---the proportion of L2 and L3 triplets---from 57.3\% to 70.5\%. The \textbf{Dense Semantic Subsumption Rate (DSSR)} of 84.0\% further indicates that most core logical facts retained by the dense baseline remain present after anchoring. Together, these results show that attention redirection improves relational depth without inducing substantial semantic loss.

\textbf{Summary and Motivation.} 
In summary, our analysis identifies the Visual Attention Sink as the principal reliability failure exposed when a monolithic graph-construction stage must localize sparse evidence, preserve global context, and extract relations simultaneously. Structural Visual Sparsity further makes full-page processing inefficient and increases susceptibility to attention drift. These findings motivate a system-level decomposition in which global grounding first establishes an explicit semantic prior and subsequent stages perform selective high-resolution extraction. The following section presents DeCoRAG, which operationalizes this mechanism-guided design.

\section{Mechanism-Guided System Design: DeCoRAG}
\label{sec:methodology}

\begin{figure*}[!t]
    \centering
    \includegraphics[width=0.85\textwidth]{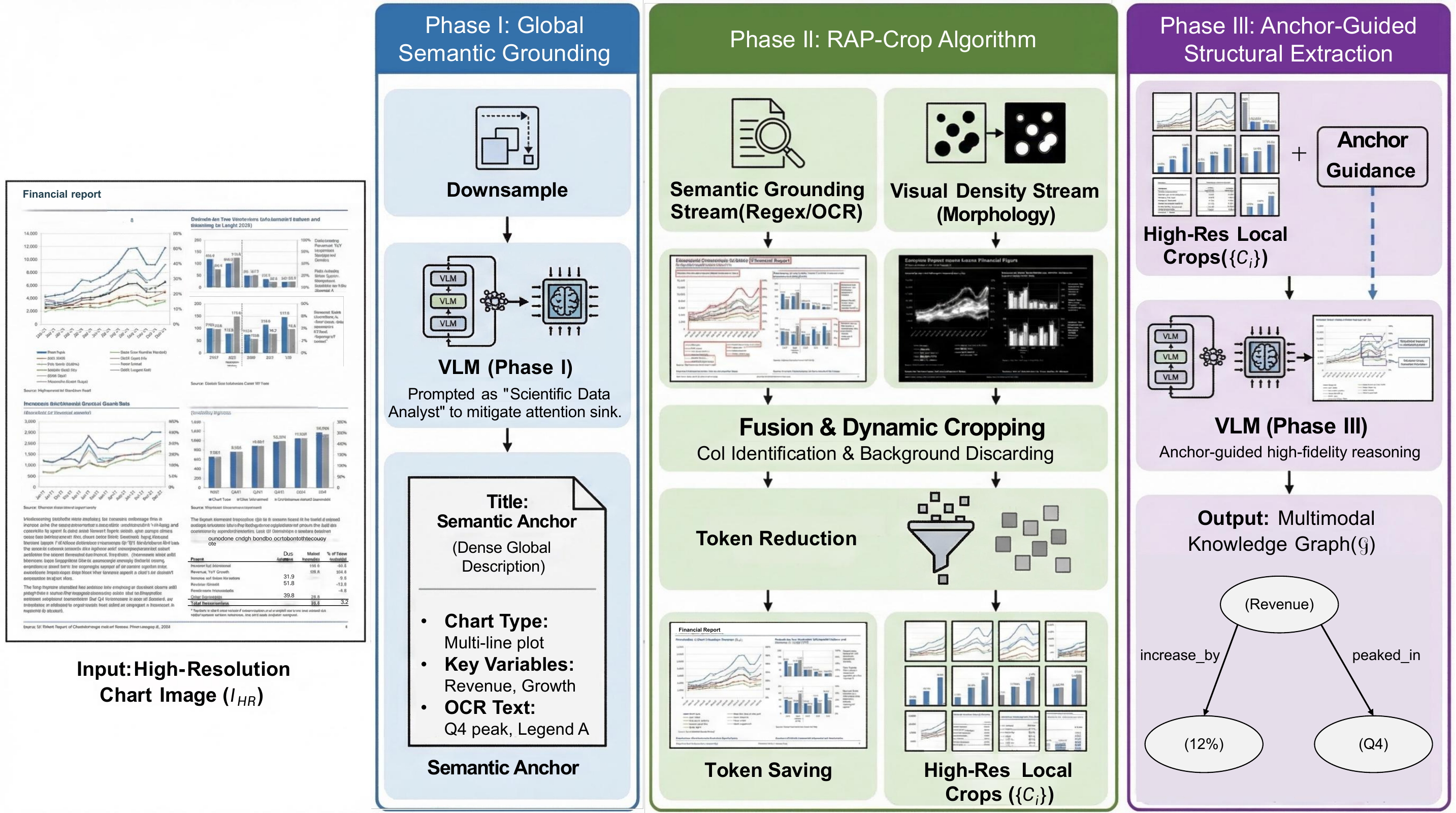}
    \caption{\textbf{The Cognitive-Decoupling Graph-Construction Subsystem within DeCoRAG.} This subsystem separates semantic planning, evidence routing, and local execution under a ``Describe-then-Extract'' paradigm. \textbf{(Phase I)} A VLM first generates a dense Global Description (Semantic Anchor) from a downsampled view, mitigating the attention sink. \textbf{(Phase II)} Guided by this semantic anchor and visual morphology, dual-stream RAP-Crop identifies Clusters-of-Interest (CoI), preserves their page coordinates, and prunes redundant background. \textbf{(Phase III)} The refined high-resolution patches, contextualized by the semantic anchor, are processed by the VLM for high-fidelity structural extraction, yielding the Multimodal Knowledge Graph used by downstream hybrid retrieval and answer generation.}
    \label{fig:architecture}
\end{figure*}

\subsection{System Overview and Execution Flow}
\label{subsec:overview}
As established in Section \ref{sec:analysis}, the unguided attention mechanism suffers from the \textbf{Visual Attention Sink}, while \textbf{Structural Visual Sparsity} renders full-page encoding computationally prohibitive. DeCoRAG is an end-to-end multimodal Graph RAG pipeline spanning offline document ingestion and graph construction, online hybrid graph retrieval, and answer generation. Within this complete pipeline, DeCoRAG addresses the diagnosed failures by replacing the monolithic graph-construction stage with a ``Describe-then-Extract'' hierarchical process (Figure \ref{fig:architecture}). Its three construction phases communicate through explicit intermediate representations---the Semantic Anchor $\mathcal{A}$, coordinate-bearing crop set $\mathcal{C}$, and Multimodal Knowledge Graph $\mathcal{G}$---before $\mathcal{G}$ is exposed to the downstream retrieval interface.

This \textbf{Cognitive Decoupling} paradigm operates in three phases: \textbf{(1) Phase I: Global Semantic Grounding.} The model generates a dense global description (Semantic Anchor) from a downsampled view, forcing attention to ground on valid semantic entities. \textbf{(2) Phase II: RAP-Crop.} Guided by this anchor and visual morphology, this algorithm dynamically crops Clusters-of-Interest, preserves their original page coordinates, and discards background noise, maximizing effective information density. \textbf{(3) Phase III: Local Structural Extraction.} The refined high-resolution patches, contextualized by the global anchor, are fed into the reasoning module to construct the final Multimodal Knowledge Graph $\mathcal{G}$ without resolution bottlenecks or attention drift.

\subsection{Phase I: Mitigating Attention Sink via Global Semantic Grounding}
\label{subsec:phase1}

As identified in Section \ref{sec:analysis}, when a VLM is directly tasked with complex structural extraction from a high-resolution image, the overwhelming visual tokens and lack of explicit spatial priors cause the attention to drift towards non-semantic boundary patches. Phase I aims to resolve this issue by explicitly establishing a strong contextual prior before any fine-grained reasoning occurs.

\textbf{Downsampled Global Input.} 
Instead of feeding the original high-resolution image $I_{HR}$, which exacerbates the quadratic complexity of visual attention, we first generate a downsampled global view $I_{global}$. This step captures the macroscopic layout and topology of the document while keeping the visual token count strictly bounded, effectively preventing an early computational bottleneck.

\textbf{Semantic Anchor Generation.} 
We instantiate the VLM with a domain-expert persona (e.g., ``Scientific Data Analyst'') and prompt it to analyze $I_{global}$. The objective is to extract a dense, information-rich summary, which we term the \textbf{Semantic Anchor} ($\mathcal{A}$). Specifically, the model is constrained to bypass superficial visual aesthetics and directly output key structural components: (1) Figure/Table Type and Semantics, (2) Structural Keypoints (e.g., Axes, Headers), (3) Key Insights, and (4) Critical OCR Text.

\textbf{Mechanistic Impact.} 
From an architectural perspective, the generation of $\mathcal{A}$ serves as a \textbf{forced attention redirection} mechanism. By requiring the model to articulate concrete visual elements (e.g., ``Revenue'', ``12\% Growth'') rather than immediately solving the complex graph extraction task, we inject strong semantic priors into the Transformer's context window. The resulting control signal increases visual grounding in subsequent reasoning and reduces the probability mass assigned to sink regions, as quantified in Section \ref{sec:analysis}.

\subsection{Phase II: RAP-Crop for Exploiting Visual Sparsity}
\label{subsec:rap_crop}

As highlighted in our mechanistic analysis (Section \ref{sec:analysis}), the \textit{Structural Visual Sparsity} of document images means that semantically valid information is highly localized, while the majority of the pixel space is redundant background. Standard sliding-window cropping inevitably slices through these semantic regions or wastes computational budget on empty patches. To resolve this, we propose the \textbf{Region-Aware Pruning and Cropping (RAP-Crop)} algorithm, a dual-stream heuristic approach designed to dynamically extract Clusters-of-Interest (CoI) while systematically discarding redundant visual tokens to minimize computational overhead (Algorithm \ref{alg:rap_crop}).

\textbf{Dual-Stream Region Detection.} 
Instead of relying on computationally heavy object detectors \cite{zhu2020deformable}, RAP-Crop fuses traditional computer vision priors with VLM-guided semantic priors through a dual-stream architecture:
\begin{itemize}
    \item \textit{Stream A: Visual Density.} We hypothesize that charting elements (lines, axes, data markers) exhibit high spatial frequency. We apply adaptive thresholding and morphological dilation on the high-resolution image $I_{HR}$ to connect localized ink pixels into contiguous blocks, yielding a set of coarse visual bounding boxes $\mathcal{B}_{vis}$.
    \item \textit{Stream B: Semantic Grounding.} To ensure critical textual elements (identified in Phase I) are not omitted, we extract a set of semantic keywords $\mathcal{K}$ from the Semantic Anchor $\mathcal{A}$ via regular expressions. We then employ a lightweight OCR engine to map these keywords back to their precise physical coordinates, producing semantic bounding boxes $\mathcal{B}_{sem}$.
\end{itemize}
The two streams play complementary roles. Stream A preserves high-recall coverage for dense visual structures even when the downsampled semantic anchor misses small text, while Stream B injects task-relevant OCR and entity cues that prevent the cropper from spending budget on visually dense but semantically irrelevant regions.

\textbf{Greedy Merging and Token Pruning.} 
The combined proposals $\mathcal{B}_{all} = \mathcal{B}_{vis} \cup \mathcal{B}_{sem}$ often contain highly overlapping or fragmented regions. We implement a greedy merging strategy that processes candidate boxes in descending order of their area and iteratively unifies adjacent boxes whose spatial distance $\text{Dist}(b, b')$ falls below a dynamic threshold $\tau=\rho\min(w,h)$, where $w$ and $h$ are the page width and height. Unless otherwise specified, $\rho=0.08$ and the crop padding ratio is 0.1. This process yields the final set of Clusters-of-Interest ($\mathcal{B}_{CoI}$). Finally, a padding ratio is applied to each bounding box within $\mathcal{B}_{CoI}$ to preserve local context. By physically cropping only these padded CoIs into high-resolution sub-image patches ($\mathcal{C}$) and \textbf{discarding the remaining background}, RAP-Crop systematically interrupts the influx of redundant tokens into the Transformer, achieving substantial token reduction without sacrificing resolution.

\begin{algorithm}[!t]
\caption{Region-Aware Pruning and Cropping (RAP-Crop)}
\label{alg:rap_crop}
\raggedright
\textbf{Input}: High-resolution image $I_{HR}$, Semantic Anchor $\mathcal{A}$, threshold ratio $\rho$, padding ratio $\alpha$\\
\textbf{Output}: High-resolution patches and original-image boxes $\mathcal{C}$
\begin{algorithmic}[1]
    \STATE $\tau \leftarrow \rho\min(\textsc{Width}(I_{HR}), \textsc{Height}(I_{HR}))$
    \STATE $I_{binary} \leftarrow \textsc{AdaptiveThreshold}(I_{HR})$
    \STATE $I_{dilated} \leftarrow \textsc{MorphologicalDilation}(I_{binary})$
    \STATE $\mathcal{B}_{vis} \leftarrow \textsc{FindContours}(I_{dilated})$
    \STATE $\mathcal{K} \leftarrow \textsc{ExtractKeywords}(\mathcal{A})$
    \STATE $\mathcal{B}_{sem} \leftarrow \textsc{LightweightOcr}(I_{HR}, \mathcal{K})$
    \STATE $\mathcal{B}_{all} \leftarrow \mathcal{B}_{vis} \cup \mathcal{B}_{sem}$
    \STATE $\mathcal{B}_{CoI} \leftarrow \emptyset$
    \WHILE{$\mathcal{B}_{all} \neq \emptyset$}
        \STATE Pop the largest box $b$ from $\mathcal{B}_{all}$
        \IF{$\exists b' \in \mathcal{B}_{all} \text{ such that } \textsc{Dist}(b, b') < \tau$}
            \STATE $b \leftarrow b \cup b'$
            \STATE Remove $b'$ from $\mathcal{B}_{all}$
        \ENDIF
        \STATE $\mathcal{B}_{CoI} \leftarrow \mathcal{B}_{CoI} \cup \{b\}$
    \ENDWHILE
    \STATE $\mathcal{C} \leftarrow \emptyset$
    \FOR{each $b \in \mathcal{B}_{CoI}$}
        \STATE $b_{pad} \leftarrow \textsc{ApplyPadding}(b, \alpha)$
        \STATE $\mathcal{C} \leftarrow \mathcal{C} \cup \{(\textsc{Crop}(I_{HR}, b_{pad}), b_{pad})\}$
    \ENDFOR
    \STATE \textbf{return} $\mathcal{C}$
\end{algorithmic}
\end{algorithm}

\subsection{Phase III: Anchor-Guided Local Structural Extraction}
\label{subsec:phase3}

With the redundant background physically eliminated by RAP-Crop and the global context encapsulated in the Semantic Anchor ($\mathcal{A}$), the framework advances to the final reasoning phase. In this stage, the VLM operates under a \textit{multi-granularity multimodal context window}, enabling it to perceive fine-grained local details without losing macroscopic topological awareness.

\textbf{Multi-Granularity Visual Prompting.} 
The input to the Phase III reasoning module is a carefully orchestrated composition. Visually, it receives the downsampled global view $I_{global}$ (to retain the overall layout) alongside the sequence of high-resolution sub-image patches $\mathcal{C} = \{c_1, c_2, \dots, c_k\}$ extracted by RAP-Crop. Textually, it is prompted with the Semantic Anchor $\mathcal{A}$ generated in Phase I, serving as a powerful cognitive prior.

\textbf{Crop-Global Coordinate Alignment.}
The downsampled global view is used only for coarse semantic and layout grounding; small-text OCR and numerical evidence are read from original-resolution crops. To preserve cross-crop topology, each crop is stored with its original page identifier and padded bounding box $[x_1,y_1,x_2,y_2]$. Triples extracted from local crops are therefore projected back into the global page coordinate system before graph merging. This design prevents a crop from becoming an isolated evidence fragment and allows relations such as containment, adjacency, and non-overlap to be recovered across crop boundaries.

\textbf{High-Fidelity Relational Reasoning.} 
By feeding $\mathcal{A}$ into the context window prior to the visual tokens, we fundamentally alter the model's attention dynamics. The VLM no longer needs to blindly search the canvas or default to the top-left visual sink tokens when faced with uncertainty. Instead, its attention is explicitly guided by the anchor to the corresponding high-resolution sub-image patches $\mathcal{C}$. This focused visual grounding allows the model to accurately read microscopic optical characters (e.g., exact axis values, footnote texts) and parse complex topological structures.

\textbf{Structured Graph Generation.} 
Ultimately, the VLM acts as a structured knowledge parser, tasked with inducing the open-schema Multimodal Knowledge Graph $\mathcal{G}$. Driven by the explicit visual grounding, the model successfully overcomes the ``Information Collapse'' failure mode identified in Section \ref{sec:analysis}. It outputs a comprehensive set of relational triples $\mathcal{G} = \{(h, r, t)\}$, where the relations $r$ transcend trivial positional prepositions (e.g., \textit{above}, \textit{left\_of}) and capture rich, logically complex dependencies (e.g., \textit{non-overlap} constraints, \textit{part\_of} hierarchies, and numerical comparisons like \textit{increase\_by}). This marks the successful recovery of the model's inherent extraction capacity, yielding a high-density, high-fidelity knowledge graph.

\subsection{System Integration: Hybrid Graph Retrieval}
\label{subsec:downstream_retrieval}
Within the complete DeCoRAG pipeline, Cognitive Decoupling redesigns the offline graph-construction subsystem while the online stage performs hybrid graph retrieval and answer generation. We integrate the constructed graph $\mathcal{G}$ with a widely adopted hybrid retrieval mechanism to instantiate this end-to-end workflow.

Specifically, given a user query, the system first employs dense vector similarity to locate semantically relevant anchor nodes within $\mathcal{G}$. Subsequently, it executes a topological traversal using Personalized PageRank (PPR) starting from these anchors to fetch the interconnected sub-graphs. These context-rich snippets are then fed into the LLM for final answer generation. The end-to-end results in Section \ref{sec:experiments} show that the higher-fidelity graphs constructed by DeCoRAG improve the evidence available to this established retrieval backend.

\section{System Evaluation}
\label{sec:experiments}

\subsection{Experimental Setup}
\label{subsec:exp_setup}

\textbf{Evaluation Scope, Datasets \& Baselines.} 
To evaluate DeCoRAG as an end-to-end multimodal Graph RAG pipeline, we examine QA reliability, component effectiveness, cross-layout generalization, offline graph-construction efficiency, and parameter robustness. We select three benchmarks exhibiting structural visual sparsity: \textbf{SPIQA} \cite{pramanick2024spiqa, masry2022chartqa, lee2023deplot, methani2020plotqa} (scientific figures/tables), \textbf{SlideVQA} \cite{tanaka2023slidevqa} (presentation slides), and \textbf{PaperTab} \cite{hui2024uda, chen2019tabfact} (academic tables). We further use \textbf{DocVQA} \cite{mathew2021docvqa} to evaluate general document VQA beyond chart- and table-centric sparsity, and \textbf{FetaTab} \cite{hui2024uda} to measure high-resolution graph-construction efficiency. We compare \textbf{DeCoRAG} against three leading technical paradigms: (1) \textbf{Standard Text-Only RAG}; (2) \textbf{End-to-End Visual Retrievers} (\textbf{ColPali-V1.2}, \textbf{ColQwen2-V1.0} \cite{wang2024qwen2vl}); and (3) \textbf{Multimodal Graph RAG} (\textbf{MMGraphRAG} \cite{wan2025mmgraphrag}).

\textbf{Evaluation Protocol.} 
Evaluating open-schema knowledge extraction is challenging as exact-match metrics penalize lexically diverse generations. Thus, we employ a 32B LLM-as-a-judge \cite{zheng2023judging, liu2023g} outputting structured rationales and scores based on Fact Consistency (40\%), Information Completeness (30\%), Logical Structure (20\%), and Expression Quality (10\%). Our primary metric is the \textbf{Semantic Pass Rate (SPR)}, where a generation ``Passes'' if its weighted score is $\ge 0.5$ (i.e., factually consistent without severe hallucinations).

\subsection{Evaluator Calibration and Evidence Recall}
\label{subsec:evaluator_calibration}

SPR measures end-to-end QA success after graph retrieval and answer generation; it is not an intermediate extraction proxy. To calibrate the LLM-as-a-judge protocol, human annotators independently judge 100 sampled answers without observing the LLM score. Table \ref{tab:spr_calibration} reports the resulting agreement as integer match rates over these 100 samples, together with evidence recall, which measures whether retrieved graph context contains the gold answer, rationale, and caption evidence. The high Human--LLM agreement on SPIQA and SlideVQA indicates that SPR is aligned with human judgment for the evaluated outputs, while evidence recall verifies that the retrieved graph snippets preserve source information used by downstream answer generation.

\begin{table}[!t]
\caption{\textbf{SPR Calibration and Evidence Recall.} Recalls are token-level/ROUGE-1 coverage of gold answer, rationale, and caption evidence in retrieved passages. SlideVQA does not provide rationale or caption fields.}
\label{tab:spr_calibration}
\centering
\scriptsize
\setlength{\tabcolsep}{2.0pt}
\begin{tabular*}{\columnwidth}{@{\extracolsep{\fill}}lcccc@{}}
\toprule
\textbf{Dataset} & \textbf{\shortstack{Human--LLM\\Agreement}} & \textbf{\shortstack{Answer\\Recall}} & \textbf{\shortstack{Rationale\\Recall}} & \textbf{\shortstack{Caption\\Recall}} \\
\midrule
SPIQA & 95\% & 68.8\% & 66.8\% & 76.8\% \\
SlideVQA & 86\% & 62.1\% & -- & -- \\
\bottomrule
\end{tabular*}
\end{table}

\subsection{Main Results: Overcoming Information Collapse}
\label{subsec:main_results}

\begin{table}[!t]
\centering
\small
\caption{\textbf{Main Results on Three High-Density Benchmarks.} Performance is measured by the Semantic Pass Rate (SPR). By employing cognitive decoupling, DeCoRAG effectively overcomes information collapse, significantly outperforming standard text-only pipelines, state-of-the-art visual retrievers, and existing multimodal graph frameworks (evaluated using the identical base VLM) across diverse document layouts.}
\label{tab:main_results}
\begin{tabular}{lccc}
\toprule
\textbf{Method} & \textbf{SPIQA} & \textbf{SlideVQA} & \textbf{PaperTab} \\
\midrule
Text-Only RAG & 78.7\% & 38.1\% & 62.1\% \\
ColPali-V1.2 & 83.5\% & 39.7\% & 47.7\% \\
ColQwen2-V1.0 & 86.8\% & 53.5\% & 66.8\% \\
MMGraphRAG & 77.8\% & 55.4\% & 63.9\% \\
\midrule
\textbf{DeCoRAG (Ours)} & \textbf{89.2\%} & \textbf{67.3\%} & \textbf{79.3\%} \\
\bottomrule
\end{tabular}
\end{table}

As shown in Table \ref{tab:main_results}, \textbf{DeCoRAG} consistently establishes a new state-of-the-art across all benchmarks, effectively overcoming ``Information Collapse''. On \textbf{SPIQA}, it achieves an 89.2\% SPR, outperforming ColQwen2 (86.8\%) and \textbf{MMGraphRAG} (77.8\%), indicating that structural graphs alone cannot compensate for the lack of visual decoupling. This superiority is amplified on structurally extreme datasets. On \textbf{SlideVQA}, which features highly irregular layouts demanding cross-region multi-hop reasoning, ColPali drops to 39.7\%, consistent with the visual-grounding failure diagnosed in Section \ref{sec:analysis}. In contrast, \textbf{DeCoRAG}, explicitly guided by its Semantic Anchor, maintains a robust 67.3\% (+27.6\% improvement). Similarly, on the extreme high-density grids of \textbf{PaperTab}, ColPali reaches 47.7\%, while our RAP-Crop-enabled system achieves 79.3\%. These results demonstrate the system-level benefit of combining explicit semantic grounding with selective high-resolution extraction across diverse layouts.

\subsection{Generalization to Conventional Document VQA}
\label{subsec:docvqa_generalization}

The main benchmarks emphasize visually sparse structural elements such as charts, tables, slides, and scientific figures. To evaluate whether the proposed cognitive decoupling remains effective in conventional document layouts, we additionally test DeCoRAG on 100 randomly sampled DocVQA queries covering scanned documents, forms, letters, and text-heavy pages. Table \ref{tab:docvqa_generalization} reports both official string-based metrics and judge-based QA metrics. DeCoRAG achieves 0.968 ANLS, 90.0\% Exact Match, and 97.0\% LLM QA Pass, indicating that the semantic-anchor and crop-grounded extraction strategy generalizes beyond sparse chart/table structures to broad document visual question answering.

\begin{table}[!t]
\caption{\textbf{Generalization Result on DocVQA.} ANLS is the official Average Normalized Levenshtein Similarity metric; LLM QA Pass follows the same end-to-end SPR criterion.}
\label{tab:docvqa_generalization}
\centering
\small
\setlength{\tabcolsep}{3.0pt}
\begin{tabular*}{\columnwidth}{@{\extracolsep{\fill}}lcccc@{}}
\toprule
\textbf{Dataset} & \textbf{ANLS} & \textbf{EM} & \textbf{\shortstack{Mean LLM\\Score}} & \textbf{\shortstack{LLM QA\\Pass}} \\
\midrule
DocVQA & \textbf{0.968} & \textbf{90.0\%} & \textbf{0.926} & \textbf{97.0\%} \\
\bottomrule
\end{tabular*}
\end{table}

\subsection{Ablation Study I: Efficacy of Semantic Anchoring}
\label{subsec:ablation_phase1}

\begin{table}[!t]
\centering
\small
\caption{\textbf{Cross-Architecture Efficacy of Semantic Anchoring.} Evaluated on SlideVQA. The consistent gains ($\Delta$) show that Phase I benefits every evaluated model scale and architecture.}
\label{tab:ablation_anchor_compact}
\begin{tabular}{lccc}
\toprule
\textbf{Model} & \textbf{w/o Phase I} & \textbf{Full} & \textbf{$\Delta$} \\
\midrule
Phi-3.5-Vis (4B)      & 47.7\% & 52.3\% & \textbf{+4.6\%} \\
InternVL2.5 (8B)      & 49.2\% & 52.7\% & \textbf{+3.5\%} \\
LLaVA-OV (7B)         & 45.4\% & 53.2\% & \textbf{+7.8\%} \\
Pixtral-12B           & 47.9\% & 53.7\% & \textbf{+5.8\%} \\
LLaVA-v1.6 (34B)      & 48.8\% & 56.4\% & \textbf{+7.6\%} \\
Qwen3-VL (32B)        & 56.8\% & \textbf{67.3\%} & \textbf{+10.5\%} \\
\bottomrule
\end{tabular}
\end{table}

To isolate the contribution of Phase I, we evaluate the system without the Semantic Anchor (\textit{w/o Phase I}) on the \textbf{SlideVQA} benchmark, a formidable stress test for reasoning across visually dense pages. Taking \textbf{Qwen3-VL (32B)} as a representative example, eliminating Phase I causes the SPR to drop from 67.3\% to 56.8\%. This degradation indicates that without a macroscopic semantic anchor, the model struggles to maintain visual grounding. As shown in Table \ref{tab:ablation_anchor_compact}, Phase I yields consistent gains across all evaluated architectures, supporting its role as a reusable system component rather than a model-specific prompt optimization.

\subsection{Ablation Study II: Dual-Stream RAP-Crop and Token Efficiency}
\label{subsec:ablation_rapcrop}

\begin{table}[!t]
\caption{\textbf{Branch-Level RAP-Crop Ablation on 50 FetaTab Queries.} Extraction tokens are graph-extraction calls after the shared semantic-anchor step; LLM QA Pass follows SPR.}
\label{tab:rapcrop_branch}
\centering
\scriptsize
\setlength{\tabcolsep}{2.5pt}
\begin{tabular*}{\columnwidth}{@{\extracolsep{\fill}}lccc@{}}
\toprule
\textbf{Variant} & \textbf{\shortstack{Extraction\\Prompt Tokens}} & \textbf{\shortstack{Extraction\\Total Tokens}} & \textbf{\shortstack{LLM QA\\Pass}} \\
\midrule
Visual-density branch only & 660K & 1.015M & 86.0\% \\
Semantic-grounding branch only & 216K & 0.322M & 84.0\% \\
Dual-stream RAP-Crop & 411K & 0.772M & 86.0\% \\
\bottomrule
\end{tabular*}
\end{table}

Table \ref{tab:rapcrop_branch} isolates the two branches of RAP-Crop on the same FetaTab subset. The visual-density branch provides high coverage and reaches 86.0\% QA Pass, but it preserves many dense regions and therefore consumes 660K extraction prompt tokens. The semantic-grounding branch is much cheaper (216K prompt tokens), but its 84.0\% QA Pass indicates that anchor-derived keywords alone can miss fine-grained visual structures. The dual-stream design keeps the visual branch's 86.0\% QA Pass while reducing extraction prompt and total tokens by 37.7\% and 24.0\%, respectively. Graph scale shows the same trade-off: semantic-only extraction yields 2.2K nodes and 3.3K edges, whereas the dual stream yields 6.2K nodes and 11.2K edges, close to the visual-only graph scale of 6.2K nodes and 11.0K edges.

\begin{table*}[!t]
\caption{\textbf{Full Offline FetaTab Graph-Construction Cost.} Similar VLM request counts indicate that RAP-Crop reduces per-request visual input rather than skipping documents.}
\label{tab:ablation_rapcrop}
\centering
\footnotesize
\setlength{\tabcolsep}{5pt}
\begin{tabular*}{\textwidth}{@{\extracolsep{\fill}}lccccc@{}}
\toprule
\textbf{Setting} & \textbf{Offline Time} & \textbf{VLM Req.} & \textbf{Prompt Tokens} & \textbf{Output Tokens} & \textbf{Total Tokens} \\
\midrule
w/o RAP-Crop & 41.85h & 124,958 & 99.54M & 36.75M & 136.29M \\
w/ RAP-Crop & 38.06h & 125,112 & 58.91M & 37.07M & 95.98M \\
Reduction & $-9.1\%$ & $+0.1\%$ & $-40.8\%$ & $+0.9\%$ & $-29.6\%$ \\
\bottomrule
\end{tabular*}
\end{table*}

To quantify RAP-Crop's ability to break the resolution-efficiency bottleneck, we measure full offline graph-construction cost on FetaTab \cite{hui2024uda}, whose high-resolution Wikipedia images emulate real-world industrial document workloads. RAP-Crop is applied only during offline graph construction; after indexing, online retrieval and answer generation follow the same downstream QA pipeline. As shown in Table \ref{tab:ablation_rapcrop}, RAP-Crop reduces prompt tokens from 99.54M to 58.91M, a 40.8\% reduction, and reduces total tokens by 29.6\%. Output tokens remain stable (+0.9\%), and VLM request counts are essentially unchanged (+0.1\%), which indicates that the gain comes from pruning redundant visual input per request rather than reducing the amount of processed evidence. The end-to-end SPR remains stable (90.3\% without RAP-Crop versus 90.6\% with RAP-Crop), showing that the observed token reduction preserves answer quality during graph construction.

\textbf{Threshold Robustness.}
RAP-Crop uses the dynamic threshold $\tau=\rho\min(w,h)$ with default $\rho=0.08$ and padding ratio 0.1. Sweeping $\rho$ over 0.06, 0.08, and 0.10 on 566 SPIQA images preserves dual-stream foreground, OCR, and reference-OCR recalls at 1.00, while the average crop count changes only from 1.18 to 1.08. This stability suggests that crop merging is not brittle across realistic layout variation.

\section{Limitations and Future Work}
\label{sec:limitations}
While robust, DeCoRAG's reasoning relies on the Phase I Semantic Anchor; severe initial hallucinations may cascade, motivating future Agentic Self-Correction mechanisms. Additionally, Stream A relies on heuristic morphological operations which may struggle with heavily degraded documents, paving the way for end-to-end differentiable routing. Finally, we plan to extend this paradigm to Inter-Page Graph Construction for multi-page industrial reasoning.

\section{Conclusion}
\label{sec:conclusion}
This paper diagnoses the \textit{Visual Attention Sink} as a model-level reliability failure systematically exposed by tightly coupled multimodal graph construction, while \textit{Structural Visual Sparsity} makes the same monolithic pipeline computationally inefficient. Controlled paired interventions show that the sink is not a simple top-left content artifact and that semantic anchoring substantially reduces both COCR and top-left attention mass. Guided by this diagnosis, \textbf{DeCoRAG} reorganizes graph construction into global semantic grounding, dual-stream evidence routing, and local high-resolution extraction. Across SPIQA, SlideVQA, PaperTab, DocVQA, and FetaTab, this mechanism-guided system improves graph-grounded QA accuracy, generalizes to conventional document VQA, and reduces offline graph-construction prompt tokens by 40.8\% without degrading answer quality. These results support pipeline-level cognitive decoupling as an accurate and cost-efficient foundation for document-scale multimodal Graph RAG.

\bibliographystyle{IEEEtran}
\bibliography{references}

\end{document}